\documentclass[a4,12pt]{article}
\usepackage{type1cm,amsmath,amssymb,color,graphics,amscd,amsfonts,mathrsfs,epsf,indentfirst,cite}
\usepackage[dvipdfm,hypertex]{hyperref}
\makeatletter

\makeatletter
\renewcommand\section{\@startsection {section}{1}{\z@}%
                                   {-3.5ex \@plus -1ex \@minus -.2ex}
                                   {2.3ex \@plus.2ex}%
                                   {\normalfont\large\bfseries}}
\renewcommand\subsection{\@startsection{subsection}{2}{\z@}%
                                     {-3.25ex\@plus -1ex \@minus -.2ex}%
                                     {1.5ex \@plus .2ex}%
                                    {\normalfont\bfseries}}

\def\btab{\begin{table}[h] \begin{center} \begin{tabular}{l lp{3in}}}
      \def\etab{\end{tabular} \end{center} \end{table}}
\def\btabm{\begin{center} \begin{tabular}}
    \def\etabm{\end{tabular} \end{center}}

\def\d{\delta}

\def\f#1#2{{\frac{#1}{#2}}}

\def\s{\sqrt}

\def\f {\frac}

\def\p{\partial}

\def\CG{{\cal G}}
\def\CH{{\cal H}}

\def\CP{{\cal P}}

\begin{document}

\begin{titlepage}
  \thispagestyle{empty}

  \begin{flushright}
  
 KUNS-2206\\
  \end{flushright}

  \vspace{2cm}

  \begin{center}
    \font\titlerm=cmr10 scaled\magstep4
    \font\titlei=cmmi10 scaled\magstep4
    \font\titleis=cmmi7 scaled\magstep4
     \centerline{\titlerm
      Ho{\v r}ava-Lifshitz Holography}

     \vspace{1.5cm}
     \noindent{{\large
         Tatsuma Nishioka
       }}\\
     \vspace{0.8cm}

    {\it Department of Physics, Kyoto University, Kyoto 606-8502,
   Japan} \\
   \vspace{0.2cm}
   {\it and} \\
   \vspace{0.2cm}
   {\it Institute for the Physics and Mathematics of the Universe, \\
 University of Tokyo, Kashiwa, Chiba 277-8582, Japan\\}

   \vspace{1cm}
   {\large \today}
  \end{center}

  \vskip 5em

  \begin{abstract}
   We derive the detailed balance
   condition as a solution to the Hamilton-Jacobi equation in the Ho{\v
   r}ava-Lifshitz gravity.
   This result leads us to
   propose the existence of the $d$-dimensional quantum field theory 
   on the future boundary of
   the $(d+1)$-dimensional Ho{\v r}ava-Lifshitz gravity from the viewpoint
   of the holographic renormalization group.
   We also obtain a Ricci flow equation of the boundary theory
   as the holographic RG flow, which is the Hamilton equation in 
   the bulk gravity, by tuning parameters in the theory.

  \end{abstract}

\end{titlepage}


A renormalizable theory of gravity, so called the Ho{\v r}ava-Lifshitz
gravity, has been attracting great interest since the advent of the seminal works of Ho{\v
r}ava \cite{Horava1,Horava2}.
This theory does not have the full diffeomorphism invariance, but 
the following anisotropic scaling with dynamical critical exponent
$z$ larger than the spatial dimensions $d$ exists
\begin{align}
 x^i &\to bx^i \quad(i=1,\dots, d)\ ,\qquad t \to b^z t \ .
\end{align}
This theory has a remarkable property such that it describes the 
non-relativistic theory of gravity in the UV regime, but becomes the Einstein 
gravity in the IR region.
It was extensively applied to a resolution of the cosmological problem 
including inflation and non-Gaussianity in
\cite{TaSo,Cal,KiKo,Mu,Br,Pi,Gao,MNTY}, and new solutions were constructed
in \cite{LMP,Nas,CoYa,CCO,CLS}.
For the other recent progress, refer the reader to \cite{Vi,MTML,CaLe,Horava3,VoWe,Je,Kl,Pal,ChHu,MyKi,CHZ,OrSu}.

It becomes convenient in the following discussion to take the Wick
rotation to the Euclidean space.
The ADM decomposition of the $(d+1)$-dimensional Euclidean space is
\begin{align}\label{ADM}
 ds^2 &= N^2dt^2 + g_{ij}(dx^i + N^idt)(dx^j + N^jdt) \ ,
\end{align}
and the extrinsic curvature for the spacelike slice is given by
\begin{align}
 K_{ij}&= \f{1}{N}(\dot g_{ij} - \nabla_i N_j - \nabla_j N_i ) \ .
\end{align}
The action of the Ho{\v r}ava-Lifshitz gravity is defined as
\cite{Horava1,Horava2}
\begin{align}\label{HLaction}
 S&= \f{2}{\kappa^2}\int^{t_0}_{-\infty}dt \int_{\Sigma_{t}}d^d
 x\s{g}N\left( K_{ij}G^{ijkl}K_{kl} +
 \f{\kappa^4}{16}E^{ij}\CG_{ijkl}E^{kl} \right) \ , 
\end{align}
where we set a future boundary at $t=t_0$ and $\CG_{ijkl}$ is the inverse of the DeWitt metric $G^{ijkl}$ of the
space of metrics
\begin{align}
 G^{ijkl}&=\f{1}{2}(g^{ik}g^{jl}+g^{il}g^{jk}) - \lambda g^{ij}g^{kl} \
 ,\\
 \CG_{ijkl}&=\f{1}{2}(g_{ik}g_{jl}+g_{il}g_{jk}) - \xi g_{ij}g_{kl} \ ,\quad
\xi= \f{\lambda}{d\lambda - 1} \ .
\end{align}
The first and second terms in (\ref{HLaction}) are the kinetic and
potential terms, respectively, and $E^{ij}$ does not contain the time
derivative of $g_{ij}$.
When we require $E^{ij}$ to be a gradient of some function $W[g]$ with
respect to the metric $g_{ij}$
\begin{align}\label{DBC}
 E^{ij}=\f{1}{\s{g}}\f{\d W[g]}{\d g_{ij}} \ ,
\end{align}
it is called a ``detailed balance condition'' \cite{Horava1,Horava2}.
In general, in the context of condensed matter physics, 
it is known that theories which satisfy the detailed balance condition have
simpler quantum properties than a generic theory, and that the
renormalization properties in $(d+1)$ dimensions are often inherited
from the simpler renormalization of the theory in $d$ dimensions with the
action $W$.
It still remains to be understood what the detailed balance
condition means in the context of the Ho{\v r}ava-Lifshitz gravity, however.
We would like to get a deep insight into a role of it.

In this paper we derive the detailed balance
condition as a solution to the Hamilton-Jacobi equation in the Ho{\v
r}ava-Lifshitz gravity.
This result leads us to
propose {\it the existence of the $d$-dimensional quantum field theory with
the effective action $W$ on the future boundary of
the $(d+1)$-dimensional Ho{\v r}ava-Lifshitz gravity} from the viewpoint
of the holographic renormalization group \cite{DVV}.
This proposal reminds us of the dS/CFT correspondence \cite{St},
while the detailed balance condition forces the cosmological constant
to be always negative\footnote{We are grateful to S. Mukohyama for pointing
out this feature.}.
In addition, we obtain a Ricci flow equation of the boundary theory
as the holographic RG flow, which is the Hamilton equation in the bulk gravity.

Although this proposal is the most salient feature of our work, we
should emphasize that we can derive the detailed balance condition
and the Ricci flow equation based on the Hamiltonian
formulation of the Ho{\v r}ava-Lifshitz gravity without the holography.

The Hamiltonian formulation in the bulk gravity affords us the renormalization
group flow of the field theory on the future boundary $t=t_0$ along the
time direction as the Hamilton equation.
Such a flow is termed a holographic renormalization group flow initiated
by \cite{DVV} (for a comprehensive review see \cite{FMS} and references therein).

The ADM decomposition (\ref{ADM}) is suitable for the Hamiltonian formulation.
The conjugate momentum associated with $g_{ij}$ in the Ho{\v
r}ava-Lifshitz action (\ref{HLaction}) is (\ $\dot{}\equiv \p/\p t$)
\begin{align}
 \pi^{ij}&= \f{1}{\s{g}}\f{\d S}{\d \dot g_{ij}} =
 G^{ijkl} K_{kl} \ ,
\end{align}
and the momenta conjugate to $N$ and $N_i$ are identically zero.
The Hamiltonian is 
\begin{align}\label{Ham}
 H 
 & = \int_{\Sigma_t}d^d x\s{g} \left[ N \CH + N_a \CP^a + 2\nabla_j(\pi^{ij}N_i)
 \right] \ , 
\end{align}
with
\begin{align}
 \CH &\equiv \pi^{ij}\CG_{ijkl}\pi^{kl} -
 \f{\kappa^2}{16}E^{ij}\CG_{ijkl}E^{kl}  \ , \\
 \CP^i &\equiv -2\nabla_j\pi^{ij} \ .
\end{align}
The last term in the second line of (\ref{Ham}) vanishes when $\Sigma_t$ is
compact space; this is the case we focus on here.

In the Hamiltonian formulation, the conjugate momentum $\pi^{ij}$
becomes the independent variable instead of $\dot g_{ij}$. 
Using the momentum, we can take the action to the one in the
first-order form
\begin{align}\label{actHam}
 S[g_{ij},\pi^{ij},N,N^i]&= \int dt \int_{\Sigma_t}
 d^dx\s{g} \left[ \pi^{ij}\dot g_{ij} - N\CH -
 N_i\CP^i \right] \ .
\end{align}
Varying this action, we obtain the Hamilton
equation 
\begin{align}\label{RGflow}
 \dot g_{ij} &= 2N \CG_{ijkl}\pi^{kl} +
 \nabla_i N_j + \nabla_j N_i \ ,
\end{align}
with the Hamiltonian and momentum constraints
\begin{align}\label{constraint}
 \CH&= \CP = 0 \ .
\end{align}

Substituting the classical solution $g_c$ into the action
(\ref{actHam}) and integrating it along the time direction,
one can express $S[g]$ as a surface integral with respect to $g_c(x,t_0)$\footnote{We are
grateful to K. Murata for discussion of this point.}
\begin{align}
 S[g=g_c]&= S_{bdy}[g_c(x,t_0)]\ . 
\end{align}
It follows from this relation that $S_{bdy}$ is the effective action of
the $d$-dimensional quantum field theory on the future boundary $\Sigma_{t_0}$
using the bulk/boundary relation $Z_{gravity}=Z_{QFT}$ and
$Z_{gravity}=\exp (-S)$ in a manner similar to the AdS/CFT correspondence \cite{Ma,GKP,Wi}.

In this case, the momentum is expressed in terms of the boundary action
(see \cite{FMS} for a careful derivation)
\begin{align}\label{MomBound}
 \pi^{ij}(x,t_0)&= \f{1}{\s{g}}\f{\d
 S_{bdy}[g_{ij}]}{\d g_{ij}}\ .
\end{align}
We use the same notation $g_{ij}$ to denote $g_c(x,t_0)$
for simplicity hereafter.
Inserting these relations into the constraints (\ref{constraint}), 
we obtain the momentum constraint
\begin{align}
 \nabla_j \left[ \f{1}{\s{g}}\f{\d S_{bdy}}{\d g_{ij}}\right] = 0 \ ,
\end{align}
which indicates the conservation law of the energy momentum tensor in the
$d$-dimensional QFT, and the Hamilton-Jacobi equation from the
Hamiltonian constraint
\begin{align}
 \left( \f{1}{\s{g}} \f{\d S_{bdy}}{\d g_{ij}}\right) \CG_{ijkl} 
\left( \f{1}{\s{g}}\f{\d S_{bdy}}{\d g_{kl}}\right)
= 
\f{\kappa^4}{16}E^{ij}\CG_{ijkl}E^{kl} \ .
\end{align}
This equation is easily solved 
\begin{align}\label{E}
 E^{ij}&= 
\f{1}{\s{g}}\f{\d W[g]}{\d g_{ij}} \ ,
\end{align}
where $W$ stands for the rescaled effective action of the $d$-dimensional QFT
\begin{align}
W[g]=\f{4}{\kappa^2}S_{bdy}[g]  \ .
\end{align}
The solution (\ref{E}) to the Hamiltonian constraint results
in the detailed balance condition (\ref{DBC}) and we find that $W$ is
the (rescaled) effective action of QFT on the future boundary of the Ho{\v r}ava-Lifshitz gravity.

It is worth mentioning that 
the Hamilton equation (\ref{RGflow})  gives us the holographic
renormalization group flow after substituting  (\ref{MomBound}) into it
\begin{align}\label{HolRG}
 \dot g_{ij}|_{t=t_0}&= \f{\kappa^2}{2}N \CG_{ijkl}\f{1}{\s{g}}\f{\d W}{\d g_{kl}} +
 \nabla_i N_j + \nabla_j N_i \ .
\end{align}
This is a simpler equation in first-order as opposed to the original equation
of motion which is second-order in time derivatives.
The Hamilton equation for $\pi^{ij}$ is automatically
satisfied and it is enough to solve (\ref{RGflow})\footnote{We are
grateful to K. Izumi for informing us of this point.}. 
As was mentioned in \cite{Horava1,Horava2}, we can easily find 
classical solutions in the Ho{\v r}ava-Lifshitz gravity by just solving
this equation.

Moreover, this equation has a remarkable property as follows.
When we take the $d$-dimensional effective action $W$ for the Einstein-Hilbert action
\begin{align}
 W&= \f{1}{\kappa_W^2}\int d^dx\s{g} (-R+\Lambda_W) \ ,
\end{align}
the $(d+1)$-dimensional theory becomes the Ho{\v r}ava-Lifshitz gravity
with dynamical critical exponent $z=2$.
The holographic RG flow (\ref{HolRG}) is given by \cite{Horava1}
\begin{align}\label{Ricciflow}
 \dot g_{ij}|_{t=t_0}&= -\f{\kappa^2}{2\kappa_W^2}N \left[ R_{ij} +
 \f{1-2\lambda}{2(d\lambda -1)}(R-2\Lambda_W)g_{ij}\right] +
 \nabla_i N_j + \nabla_j N_i \ .
\end{align}
If we take $\lambda =1/2$, $\kappa_W=\kappa/2$, $N=1$ and
 $N_i=0$, this 
becomes the Ricci flow equation\footnote{We can set $N_i=0$ using the
 $d$-dimensional diffeomorphism, but we are not sure if we can take
 $N=1$ by the reparametrization of the time $t\to f(t)$.}.
From this viewpoint one may say that {\it the Ricci flow in $d$ dimensions
is the holographic RG flow to the $(d+1)$-dimensional Ho{\v
 r}ava-Lifshitz gravity with $z=2$ and $\lambda=1/2$}.


One simple but interesting application of our results is that 
static solutions in the Ho{\v r}ava-Lifshitz gravity are obtainable by
solving\footnote{One can find the same discussion in \cite{Horava1}
under the assumption of (\ref{HolRG}).}
\begin{align}\label{dW}
 \f{\d W[g]}{\d g_{ij}}&= 0 \ .
\end{align}
In the case of the four-dimensional Ho{\v r}ava-Lifshitz gravity with
$z=3$, $W$ is the Einstein-Hilbert action with the gravitational
Chern-Simons term in three dimensions ({\it i.e.}\,the topologically massive gravity).
(\ref{dW}) is just the Einstein equation in TMG and the interesting
solutions were constructed in \cite{Nu,BoCl,ALPSS}.
For example, we can construct the four-dimensional solitonic solution by
the use of the Euclidean warped AdS$_3$ black hole.
It would be of wide interest to investigate such a solution in higher
dimensions with different dynamical exponent $z$.

 \vspace{0.5cm}
 \centerline{\bf Acknowledgements}
 We are grateful to 
K. Izumi, S. Minakami,
 S. Mukohyama, K. Murata, T. Kobayashi for valuable discussions, and
 S. Horiuchi for careful reading of this manuscript.
 This work is supported by JSPS Grant-in-Aid for Scientific Research
 No.\,19$\cdot$3589.



\end{document}